\documentclass{PoS}
\usepackage{upgreek}

\newcommand{\mus}{\,\ensuremath{\mathrm{\upmu s}}}
\newcommand{\ns}{\,\ensuremath{\mathrm{ns}}}
\newcommand{\nm}{\,\ensuremath{\mathrm{nm}}}

\title{Paving the way for fundamental physics tests with singly-ionized helium}

\ShortTitle{Fundamental physics with He$^+$}

\author{\speaker{J.J.~Krauth}, %\thanks{A footnote may follow.}
  L.S.~Dreissen, C.~Roth, E.L.~Gr\"undeman, M.~Collombon, M.~Favier,\newline and K.S.E.~Eikema\\
        LaserLaB, Department of Physics and Astronomy, Vrije Universiteit, De Boelelaan 1081,\newline 1081 HV Amsterdam, The Netherlands\\
        E-mail: \email{j.krauth@vu.nl}, \email{k.s.e.eikema@vu.nl}}

\abstract{High-precision laser spectroscopy of atomic hydrogen has led to an impressive accuracy in tests of bound-state quantum electrodynamics (QED). At the current level of accuracy many systematics have to be studied very carefully and only independent measurements provide the ultimate cross-check. This has been proven recently by measurements in muonic hydrogen, eventually leading to a significant shift of the CODATA recommended values of the proton charge radius and the Rydberg constant. 
We aim to contribute to tests of fundamental physics by measuring the 1S-2S transition in the He$^+$ ion for the first time. Combined with measurements in muonic helium ions this can probe the value of the Rydberg constant, test higher-order QED terms, or set benchmarks for ab initio nuclear polarizability calculations. We extend the Ramsey-comb spectroscopy method to the XUV using high-harmonic generation in order to excite a single, trapped He$^+$ ion.}

\FullConference{International Conference on Precision Physics and Fundamental Physical Constants - FFK2019\\
		9-14 June, 2019\\
		Tihany, Hungary}

\begin{document}

\section{Introduction}
\label{sec:intro}

Simple atomic systems are the ideal probe to test fundamental physics. The most prominent example is the hydrogen atom. Its 1S-2S transition energy, for example, can nowadays be calculated with an impressive relative accuracy of below 10$^{-12}$ \cite{Yerokhin:2019:LS,Karshenboim:2019:1SLS}. Thanks to great progress in the field of laser spectroscopy over the past decades, the latest measurements of this transition have reached an even higher relative accuracy of $4\times10^{-15}$ \cite{Parthey:2011:PRL_H1S2S}. The theory prediction of the 1S-2S transition relies on fundamental constants such as the Rydberg constant, which have to be measured experimentally. A common choice to measure the Rydberg constant and test theory is the combination of a measurement of the 1S-2S transition and a transition from the 2S state to a higher-lying $n$S/D state, e.g.\ the 8S or 12D state. With increasing accuracy this test was hampered by the uncertainty in the value of the proton charge radius and therefore additional measurements in other systems were required. In 2010, the CREMA collaboration measured the proton charge radius in muonic hydrogen \cite{Pohl:2010:Nature_mup1}. This value improved the accuracy of the previous best value by one order of magnitude, but at the same time created a discrepancy of about six standard deviations with the by then recommended CODATA value. This discrepancy is known as the proton radius puzzle \cite{Pohl:2013:ARNPS,Carlson:2015:Puzzle}. With the latest hydrogen spectroscopy measurements \cite{Beyer:2017:2S4P,Fleurbaey:2018:1s3s,Bezginov:2019:LS}, included in the CODATA 2018 report, this puzzle seems to reach a solution. Using the value of the proton radius from muonic hydrogen, the limitations for testing bound-state quantum electrodynamics (QED) in atomic hydrogen now arise from calculations of specific QED contributions. More precisely, from calculations of radiative recoil corrections and two- and three-loop contributions which include the $B_{60}$, and $C_{50}$ coefficients \cite{Mohr:2016:CODATA14,Karshenboim:2019:1SLS}. The latter two scale with large powers of the nuclear charge $Z$, and are therefore strongly enhanced in hydrogen-like helium.

Apart from a simple crosscheck with the hydrogen measurements, a measurement of the 1S-2S transition in hydrogen-like helium has the potential to also improve our understanding of fundamental physics on a deeper level. In order to specify what can be learned from a 1S-2S measurement in He$^+$ we need to look at the current limitations for the prediction of the transition energy. A detailed but slighly dated review is found in Ref.\,\cite{Herrmann:2009:He1S2S}. In this part we provide a brief update, including the most recent developments. Following the notation of \cite{Pohl:2017:DSpec}, the energy levels for hydrogen-like systems can be described in a simplified way by
\begin{equation}\label{eq:levels}
  E(n,l,j)/h = -\frac{Z^2 c R_\infty}{n^2} \frac{m_{red}}{m_e} + \frac{E_{NS}}{n^3}\delta_{l0} + \Delta(n,l,j)
\end{equation}
where the first term is the classical Bohr term, limited by the uncertainty in the value of the Rydberg constant $R_\infty$, the second term is the finite nuclear size term, limited by the uncertainty in the value of the nuclear charge radius (in leading order only existent for $S$-levels), and the last term contains everything which was not included up to this point, such as the fine and hyperfine structure, relativistic effects and QED. It is noteworthy that the finite size effect $E_{NS}$ and above mentioned limiting QED contributions scale with $Z^4$ and $Z^{6+}$, respectively. A nuclear charge of $Z=2$ therefore enhances the sensitivity to the above mentioned limiting QED contributions compared to hydrogen. The limitation can be quantified by including the available values for $R_\infty$ and the nuclear size of the alpha particle $r_\alpha$ which we discuss in the following part.

The most precise value for the Rydberg constant is obtained by combining the 1S-2S measurement from Ref.\,\cite{Parthey:2011:PRL_H1S2S} with the Lamb shift measurement from muonic hydrogen \cite{Pohl:2010:Nature_mup1,Antognini:2013:Science_mup2}. This results in $R_\infty=3.289\,841\,960\,249\,5(10)(25)\times10^{15}\,\mathrm{Hz}/c$, with an uncertainty from the proton charge radius of $1\,\mathrm{kHz}/c$ and from QED of $2.5\,\mathrm{kHz}/c$. The most recent recommended value from the CODATA 2018 adjustment is  $R_\infty=3.289\,841\,960\,250\,8(64)\times10^{15}\,\mathrm{Hz}/c$ \cite{Mohr:2020:CODATA18}. These two values of $R_\infty$ lead to an uncertainty in the Bohr term for the 1S-2S transition frequency in He$^+$ of $9\,\mathrm{kHz}$ and $19\,\mathrm{kHz}$, respectively.

The latest published value of the charge radius of the alpha particle is obtained from electron scattering with a value of $r_\alpha = 1.681(4)\,\mathrm{fm}$ \cite{Sick:2008:rad_4He}. However, the CREMA collaboration has performed a more precise measurement of $r_\alpha$ in muonic helium which is expected to be published soon. Since the nuclear charge radius extracted from muonic helium will be strongly limited by theory (mainly due to the nuclear polarizability contribution), we can use the theory uncertainty of $r_\alpha$ of 0.0008\,fm \cite{Diepold:2016:mu4HeTheo} to estimate the effect on the uncertainty of the 1S-2S transition frequency in He$^+$, which then yields an uncertainty of about $60\,\mathrm{kHz}$. An improvement on the calculation of the nuclear polarizability in muonic helium will result in a direct reduction of this uncertainty.

Recently there has also been impressive progress in QED calculations \cite{Yerokhin:2019:LS,Karshenboim:2019:1SLS}. The largest uncertainty arises from the two-loop contribution, more precisely from the $B_{60}$ coefficient. The recommended CODATA 2014 value of this coefficient limited the prediction of the 1S-2S transition frequency in He$^+$ to 110\,kHz \footnote{This number can be obtained using the respective coefficient given in Tab.\,IX of the CODATA report \cite{Mohr:2016:CODATA14} and Eq.\,(49) therein.}. The next limiting terms are the three-loop contribution (the limiting coefficient is $C_{50}$) as well as radiative-recoil corrections. With the very recent work from the groups of Pachucki and Karshenboim, these uncertainties could be reduced significantly, such that by now the combined uncertainty of the three contributions mentioned above in the He$^+$ 1S-2S transition amounts to roughly 41\,kHz \cite{Karshenboim:2019:1SLS}, which is therefore slightly smaller than the uncertainty arising from the finite size effect (60\,kHz).

From the discussed contributions we conclude the following: Given the above uncertainties it does not seem to be useful to extract a Rydberg constant from He$^+$ since its accuracy will not be able to compete with the one obtained from hydrogen. Instead we can test the mentioned QED contributions to an unprecedented level, limited by the uncertainty in the finite size effect. 
Assuming that QED is correct within the given uncertainty, we can extract the alpha particle charge radius. The accuracy of that value will then be limited by the QED calculations. A comparison with the measurement of the Lamb shift in muonic helium will yield a prediction for the polarizability of the helium nucleus. This prediction will serve as a benchmark for nuclear polarizability calculations.

\section{From Ramsey's molecular beam method to Ramsey-comb spectroscopy in the XUV spectral range}
\label{sec:rcs}

The energy needed to excite the 1S-2S transition in He$^+$ corresponds to the energy of a 30\nm{} photon, which is in the extreme ultraviolet (XUV) range.

To the best of our knowledge there are currently two groups working towards a measurement of the He$^+$ 1S-2S transition, which pursue different methods. One is the group of Th.\ Udem \cite{Herrmann:2009:He1S2S} at the Max-Planck Institute of Quantum Optics (MPQ) in Garching, the other one is the group of K.S.E. Eikema at the Vrije Universiteit Amsterdam. The group at MPQ applies pure direct frequency comb (FC) spectroscopy. By means of intra-cavity high-harmonic generation (HHG) they reach the required peak powers in the fundamental for efficient upconversion to a wavelength of 60\nm{}. The 1S-2S transition is then excited with two counterpropagating 60\nm{} photons. This method requires the entire FC to be amplified and upconverted which is very challenging, but at the same time keeps the narrow-band structure of the FC laser.

We aim to apply a different method, called Ramsey-comb spectroscopy to excite the 1S-2S transition. This method has been developed and refined in our lab over the last years as shortly summarized here. We transfer Ramsey's measurement of separated oscillatory fields in the microwave regime \cite{Ramsey:1950:MBR} into the optical ultraviolet regime using two amplified and upconverted pulses out of the infinite pulse train of a FC laser. Instead of separating the interactions of the two pulses with the atoms in space, the interaction is now separated in time. This is possible due to the well controlled phase and time relation between the FC pulses.

\begin{figure}
  \centering
  \includegraphics[width=0.9\textwidth]{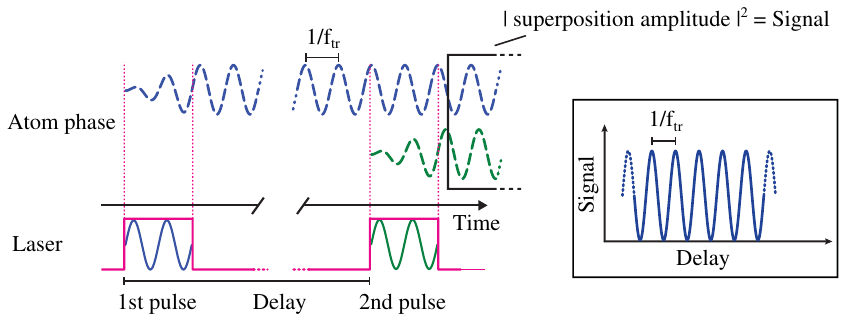}
  \caption{The principle of Ramsey spectroscopy \cite{Morgenweg:2014:PRA} in the time domain. A first resonant laser pulse creates a superposition of the atomic ground and excited state. The phase of this state evolves with the transition frequency according to $2\pi f_\mathrm{tr}\,t$. The second pulse does the same, with a very well defined delay and phase. A Ramsey fringe is obtained by scanning the interpulse delay on an attosecond level, recording the excited state population which is proportional to the squared amplitude of the interference of the two superposition states. This is usually done via state-selective ionization.}
  \label{fig:ramsey}
\end{figure}
The principle of the measurement is shown in Fig.\,\ref{fig:ramsey}. The first resonant pulse puts the atomic system into a superposition state of ground and excited level. The phase of this state evolves proportionally with the transition frequency according to $2\pi f_\mathrm{tr}\,t$. The second pulse creates another superposition state which interferes with the first one. Depending on the precise time delay and optical phase between the pulses this interference is either constructive or destructive. The transition frequency is therefore measured by scanning the delay (phase would be possible too) between the two pulses on the attosecond scale (by changing the repetition rate $f_\mathrm{rep}$ of the FC). The signal is a Ramsey fringe which can be obtained using e.g.\ state-selective ionization and a time-of-flight setup \cite{Altmann:2016:Kr}, but also nondestructive methods exist which we will apply for He$^+$ (Sec.\,\ref{sec:helium}). The signal is proportional to
\begin{equation}
  \label{eq:signal}
  S \propto \cos(2\pi f_\mathrm{tr} \Delta t - \Delta\phi)
\end{equation}
where $f_\mathrm{tr}$ is the transition frequency, $\Delta t$ the interpulse delay and $\Delta\phi$ a systematic phase shift which has to be characterized. It includes the carrier envelope offset phase, which is known, and the phase shift induced by amplification and e.g. upconversion or light shift. These phase shifts have to be characterized very carefully.

We demonstrated a Ramsey measurement in the deep ultraviolet (DUV) regime in 2005 in krypton \cite{Witte:2005:DUV_ramsey}. By means of a pair of amplified (using a parametric amplifier) and upconverted subsequent pulses from a frequency comb, a two photon transition ($2\times 212.5\,$nm) was measured with a relative precision of $1.2\times10^{-9}$.

%\begin{figure}
%  \centering
%  \includegraphics[width=0.7\textwidth]{figures/fringe}
%  \caption{Ramsey fringe recorded from a Ramsey measurement in the helium atom in 2010 \cite{Kandula:2010:XUV_comb_metrology} as a function of the repetition rate $f_\mathrm{rep}$ of the FC laser. The measured transition is a one photon transition at 51.5\nm{}. The signal (blue dots) is obtained via state-selective ionization of the helium atoms and a time-of-flight measurement. The data is fitted with a sinusoidal curve (red line).}
%  \label{fig:fringe}
%\end{figure}
In 2010, we extended the Ramsey spectroscopy method to the XUV range by combining it with HHG to measure a single-photon transition in the $^4$He atom \cite{Kandula:2010:XUV_comb_metrology} at 51\nm{}. This was the first absolute frequency measurement in the XUV wavelength range (below 100\nm{}). The Ramsey measurements done so far were limited by uncertainties in the systematic phase shift $\Delta\phi$ (see Eq.\,\ref{eq:signal}) which had to be measured and characterized very carefully. Furthermore, it was only possible to select subsequent pulses ($\sim10\,$ns delay), whereas a longer pulse delay would linearly increase the precision.

These limitations were overcome in 2013, with the development of the Ramsey-\textit{comb} spectroscopy (RCS) method, demonstrated in Ref.\,\cite{Morgenweg:2013:RCS} in the near-infrared. The RCS method is a combination of Ramsey measurements at different so-called macro-delays (given by a multiple integer $n$ of the repetition time $\mathrm{T_{rep}}$ of the FC). For each macro-delay $n\mathrm{T_{rep}}$ a Ramsey fringe is recorded, leading to a series of Ramsey fringes spaced by the repetition time of the comb. The transition frequency $f_\mathrm{tr}$ results from the relative phase difference between these fringes. In this way, the influence of all delay-independent phase shift contributions to $\Delta\phi$ cancel, such as those induced by the parametric amplification process or the AC-Stark effect (an effect on the wavefunctions, which takes the form of a constant phase shift after the interaction with the two pulses). Furthermore interpulse delay times of several 100\ns{} were achieved and can further be increased to the microsecond level, leading to much higher accuracies.
With the RCS method also several transitions at the same time can be measured as described in Ref.\,\cite{Morgenweg:2014:PRA}.

The Ramsey-comb method in the DUV regime was demonstrated for the first time with a measurement in krypton in 2016 \cite{Altmann:2016:Kr}, and later of the EF-X(Q1) transition in H$_2$ \cite{Altmann:2018:H2}. We improved the accuracy of this H$_2$ transition by two orders of magnitude, enabling a more precise determination of the dissociation energy $D_0$, a benchmark quantity for molecular quantum theory.

Paving the way for Ramsey-comb measurements in the XUV, we recently demonstrated that Ramsey-comb spectroscopy can be successfully combined with high-harmonic generation with a measurement of a single photon transition at 110\nm{} in xenon \cite{Dreissen:2019:Xe}. Using the xenon atom as a phase detector, we find that phase shifts induced from ionization during the HHG process can be reduced to a negligible level. We determined the absolute transition frequency, improving the previous best value by four orders of magnitude \cite{Yoshino:1985:xenon}. The relative precision of this measurement is $2.3\times10^{-10}$, which is 3.6 times better than the previous most accurate measurement with light from a HHG source \cite{Cingoz:2012:directFCS}. Our measurement accuracy is limited by transit-time broadening, the Doppler effect, and the $\sim22\,$ns excited state lifetime. These limitations are absent for He$^+$, which has a very long (1.9\,ms) upper state lifetime, and in addition can be trapped and sympathetically cooled.  

\section{The He$^+$ experiment}
\label{sec:helium}

\begin{figure}
  \centering
  \includegraphics[width=0.82\textwidth]{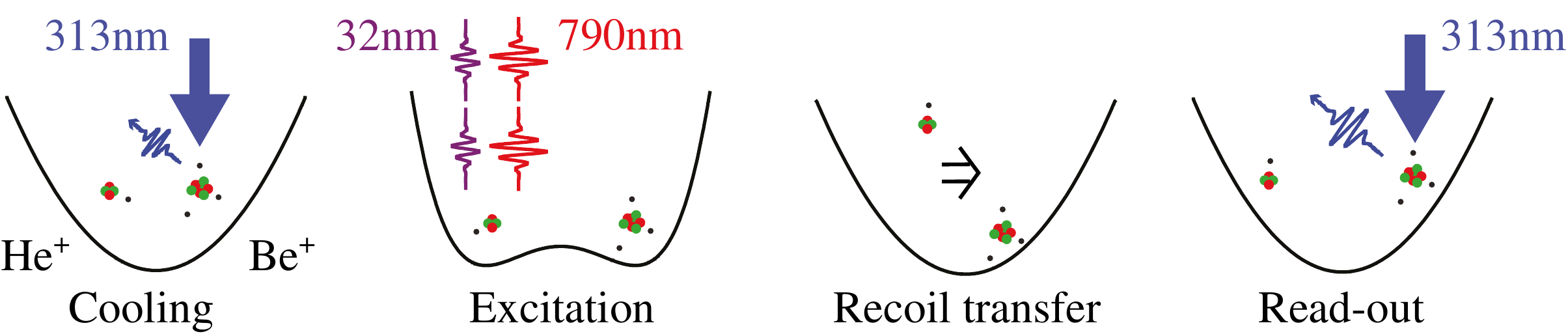}
  \caption{The basic principle of the He$^+$ 1S-2S experiment. The He$^+$ ion is sympathetically laser cooled with a Be$^+$ ion to the ground state of motion. For the actual Ramsey-comb measurement the ions are separated with a double well potential. The absorption of the XUV photon heats the He$^+$ ion by more than 10 phonons. These motional quanta are transferred to combined modes with the Be$^+$ ion when they are brought in close proximity again. In the last step the Be$^+$ ion is read out via state-dependent fluorescence.}
  \label{fig:cycle}
\end{figure}

We aim to excite the 1S-2S transition of a single trapped and cooled helium ion with two wavelength unequal, co-propagating photons (the fundamental at 790\nm{} + the 25\textsuperscript{th} harmonic at 32\nm{}) by means of the Ramsey-comb spectroscopy method. Instead of using a Doppler-free configuration (counterpropagating equal photons), the different approach of unequal photons enables us to increase the transition probability in two ways at the same time. First, we can use the high pulse power available in the fundamental beam and second, the virtual intermediate level comes much closer to the 2P level. The expected Doppler shift which usually arises in this unequal photon configuration can still be canceled as discussed later. The basic idea of the measurement is sketched in Fig.\,\ref{fig:cycle}.
The He$^+$ ion will be trapped in a Paul trap and sympathetically cooled to the ground state of motion with a single laser cooled (313\,nm) Be$^+$ ion. After the cooling process, the two ions are separated with a double well potential and the helium ion is interrogated with the Ramsey-comb laser. In case of a successful internal (electronic) excitation to the 2S state, the helium ion will also be externally excited by the large recoil of the absorbed XUV photon. In the next step, the helium and the beryllium ion are put close together again and the motional quanta from the helium ion are coupled to the collective modes which He$^+$ has together with the Be$^+$ ion. The Be$^+$ ion is then read out via state-dependent fluorescence. This scheme is similar to the so-called quantum logic type scheme as e.g.\ applied for single ion optical clocks \cite{Schmidt:2005:quantum_logic}.
In this way the 1S-2S transition in He$^+$ can be detected in a non-destructive manner (until multi-photon ionization leads to formation of He$^{2+}$, which we estimate happens after a few hundred excitation detections). Depending on the different He$^+$ ion loss mechanisms, we can tune the 790\,nm power for the best compromise between excitation rate (which can be near 100\% at high 790\,nm power) and double-ionization of He$^+$.

\begin{figure}
  \centering
  \includegraphics[width=0.76\textwidth]{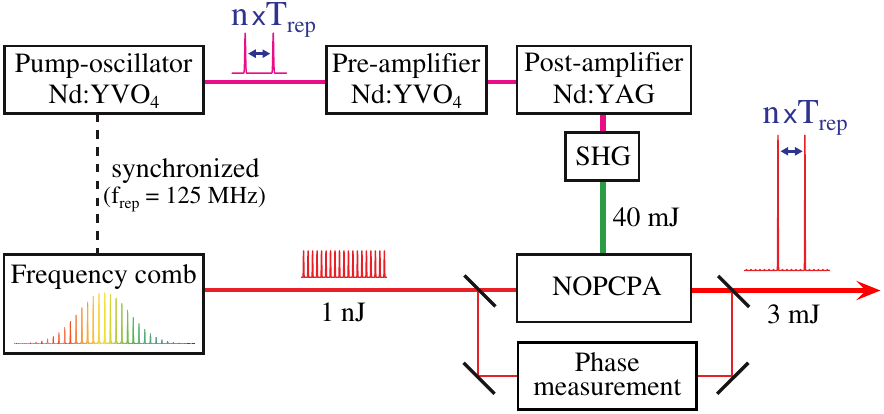}
  \caption{Sketch of the Ramsey-comb laser system. The pulse pairs needed for the Ramsey-comb spectroscopy are selected from an infinite pulse train created in a FC laser via amplification in the noncollinear optical parametric chirped pulse amplifier (NOPCPA). The pump pulses for the NOPCPA are created in the Nd:YVO$_4$ oscillator whose repetition rate is synchronized to the FC laser. After picking the desired pulse pair from the Nd:YVO$_4$ laser with fast modulators, the pulses are amplified in two stages. The expected pump pulse energy after doubling to 532\,nm is 40\,mJ, which is used to pump the 3-stage NOPCPA. After the NOPCPA, the amplified Ramsey-comb pulse pairs are then send to the vacuum chamber shown in Fig.\,\ref{fig:vacuum}.}
  \label{fig:laser}
\end{figure}
\begin{figure}
  \centering
  \includegraphics[width=\textwidth]{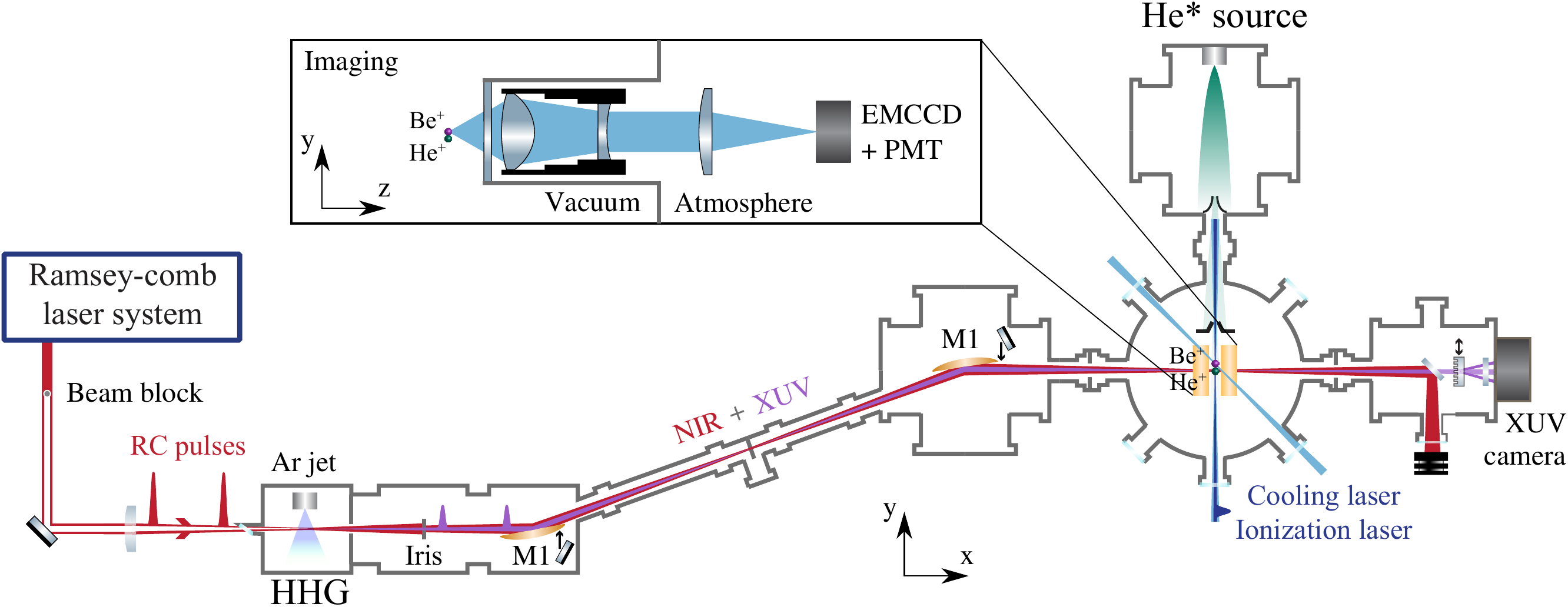}
  \caption{The vacuum setup for the He$^+$ experiment. The Ramsey pulses enter the setup from the right side where they are focused in an Ar jet for high-harmonic generation (HHG). The focus is imaged via two toroidal mirrors (M1 and M2) at grazing incidence onto He$^+$ ion in the spectroscopy segment of the ion trap. The inset shows the imaging system which is on top of the measurement chamber (note the coordinate system indicators).}
  \label{fig:vacuum}
\end{figure}

The excitation laser system is sketched in Fig.\,\ref{fig:laser}. The excitation pulses are created in an ultra-low phase noise (ULN) Er:fiber frequency comb from Menlo systems (locked to an ultra-stable (< 1\,Hz) linewidth reference laser at 1542\,nm) with a frequency doubled output at 790\nm{}. We selectively amplify two pulses out of the FC pulse train in a noncollinear optical parametric chirped-pulse amplifier (NOPCPA, in the following simply OPA) \cite{Witte:2012:OPA} to a few mJ per pulse. The pump laser for the OPA is similar to the one previously used \cite{Galtier:2017:DUV_RCS} for the measurements in krypton, hydrogen, and xenon. The pump pulses are created in a Nd:YVO$_4$ oscillator which is synchronized with the frequency comb. After picking two pulses with the desired interpulse delay from this oscillator, they are amplified in two stages. After frequency doubling to 532\nm{} the pump pulses have a pulse energy of about 40\,mJ and are overlapped in the OPA with the seed pulses from the FC. In contrast to typical laser amplifiers via inversion, the OPA does not have a memory effect. This means we can amplify selected FC pulses without any effect on the second pulse. The phase shifts induced in the OPA are measured by means of a spectral interferometer. After amplification the Ramsey pulses enter the vacuum setup shown in Fig.\,\ref{fig:vacuum}. This setup has been designed specifically for the He$^+$ measurement and was successfully tested with the measurements in xenon. In the first part of the chamber the Ramsey pulses are focused in an argon jet to create harmonics up to 25\textsuperscript{th} order. The focus at the HHG interaction zone is then imaged via two toroidal mirrors at grazing incidence to the center of the ion trap in the measurement chamber. The imaging system to observe the fluorescence contains a self-build infinity-corrected objective including a biaspheric lens to maximize the photon collection efficiency and allow diffraction-limited imaging.

We will use a modified version of an ion trap developed by our collaborators at PTB in Braunschweig \cite{Herschbach:2012:IonTrap,Pyka:2014:IonTrap}. The design is a segmented Paul trap, optimized to minimize micromotion and heating rates.
In order to load the trap we apply a metastable helium source and a beryllium ablation target. Both species will be photoionized by a picosecond pulse at 355\,nm when passing the center axis of the loading segment. After Doppler cooling, both ions are shuttled to the spectroscopy segment where Raman sideband cooling is applied on the Be$^+$ ion for all three directions of motion. The next step is the actual excitation process and the previously mentioned readout.

As any other high-precision measurement, our method is influenced by systematic effects, and we briefly list the important ones below.

As long as the ion is cooled to the motional ground state, Doppler effects are absent. However, with the absorption of the XUV photon the He$^+$ ion experiences a large recoil. This means that the excited part of the superposition state does move and therefore can induce phase shifts. We circumvent this problem by synchronizing the pulses with the ion motion. In this way we ensure that the excited state wave packet is always overlapping with the ground state wave packet when the second Ramsey pulse arrives. A relative synchronization accuracy of below 10$^{-4}$ has been achieved before \cite{Johnson:2016:RFstabilization} and will reduce the systematic shift to the desired level.

Since the control over the phase of the excitation pulses is essential for our method, we have to take a closer look at the sources of phase noise in the setup. With increasing phase noise we lose contrast in our Ramsey signal. The first source of phase noise is the FC laser itself, which has an integrated phase noise of about 30\,mrad between 100\,Hz and 2\,MHz, measured at 1560\nm{}. This phase noise is then multiplied by a factor of 52, due to the upconversion of 1560\nm{} light to a wavelength of 30\nm{}. This yields an integrated phase noise of 1.56\,rad for long interpulse delays (above 1\mus{} / below 1\,MHz) which corresponds to a contrast reduction in the signal to about 30\,\% just from the FC phase noise.
However, for our initial experiments the relevant noise frequency range is above 2.5\,MHz, i.e. for interpulse delays up to 400\ns{}, where the phase noise is almost negligible. For later measurements with interpulse delays of several microseconds or more we will reduce the phase noise of the FC further by using an external EOM.\\
A second source of phase noise arises from the OPA. In the current RCS setup for the measurements in xenon we measured a phase noise from the OPA of 50-70\,mrad for a fundamental of 770\nm{}. However, for the He$^+$ experiment we are now setting up an improved OPA version where phase noise will be reduced to a level of about 15\,mrad at 790\nm{}, well below the FC phase noise. With our experiment in xenon we have demonstrated that the HHG process itself actually induces very little phase noise in the harmonic light, therefore this contribution does not significantly add to the noise generated by the FC laser or the OPA. 

Due to the high intensity of the 790\nm{} beam at the He$^+$ ion, the energy levels experience a strong light shift. The effect on our measurement, however, completely cancels for the Ramsey-comb spectroscopy method \cite{Morgenweg:2014:PRA,Altmann:2018:H2}, where only the relative phase between pulse pairs with different delays is measured. This means that a precise control over the pulse energy at the level of 0.1\% of the Ramsey pulses is required, which we have already demonstrated in previous measurements.\\

In conclusion, the measurement of the 1S-2S transition of He$^+$ will provide an independent cross check of the fundamental physics tests performed in atomic hydrogen. Using the Rydberg constant obtained from measurements in H one can set a constraint on the sum of the higher-order two-loop QED contributions and the finite size effect, which contribute similar uncertainties. We will do the measurement by means of the Ramsey-comb spectroscopy method developed in our lab, and are now in the final phase of setting up the experiment. With a measurement in xenon we have tested the vacuum setup and successfully demonstrated the combination of Ramsey-comb spectroscopy and high-harmonic generation for the first time. Furthermore, we could show that phase shifts due to HHG process can be avoided for He$^+$.  The first test measurements with He$^+$ are planned for the coming year. Our initial goal is to measure the transition frequency with an accuracy of about 1\,kHz, corresponding to a relative accuracy of $1\times10^{-13}$, and with adjustments to the laser system we envision possibilities to improve on this with one or two orders of magnitude.

K.S.E. Eikema acknowledges the European Research Council for an ERC-Advanced grant under the European Union's Horizon 2020 research and innovation programme (Grant Agreement No. 695677) and FOM/ NWO for a Program grant (16MYSTP).

\bibliographystyle{mysty1}
\bibliography{ref}

%\begin{thebibliography}{99}
%\bibitem{...}
%....
%
%\end{thebibliography}

\end{document}